\documentclass[aps,pra,floatfix,onecolumn,groupedaddress,superscriptaddress,nofootinbib,notitlepage,amsmath,amssymb]{revtex4-1}
\usepackage{times,graphics,graphicx,bm,bbm,bbold,amsfonts,mathtools,dsfont,float,hyperref,amsthm}
\hypersetup{colorlinks,linkcolor={blue},citecolor={blue},urlcolor= {blue}}
\usepackage{mathtools}

\newsavebox{\mstrut}
\newcommand{\bbra}[1]{%
	\sbox{\mstrut}{\(#1\)}%
	\mathinner{\left\langle\kern-0.3\ht\mstrut\left\langle{#1}\right|\mkern-2mu\right|}%
}
\newcommand{\kett}[1]{%
	\sbox{\mstrut}{\(#1\)}%
	\mathinner{\left|\mkern-2mu\left|{#1}\right\rangle\kern-0.3\ht\mstrut\right\rangle}%
}

\newcommand\cysout{\bgroup\markoverwith{\textcolor{cyan}{\rule[0.5ex]{2pt}{1.2pt}}}\ULon}
\newcommand\osout{\bgroup\markoverwith{\textcolor{orange}{\rule[0.5ex]{2pt}{1.2pt}}}\ULon}

\newtheorem{remark}{Remark}

\def\be{\begin{equation}}
	\def\ee{\end{equation}}
\def\ba{\begin{eqnarray}}
	\def\ea{\end{eqnarray}}
\def\la{\langle}
\def\ra{\rangle}

\def\h{\hskip 1cm}

\def\lo{\longrightarrow}
\def\a{\alpha}

\begin{document}
	
	\title{Witnessing entanglement of remote particles with incomplete teleportation}
	
	\author{Vahid Jannessary\footnote{email: vahid.jannesary@gmail.com}}
	\affiliation{Department of Physics, Sharif University of Technology, Tehran 14588, Iran}

	\author{ Fatemeh  Rezazadeh \footnote{email: sfrezazadeh@yahoo.com}}
	\affiliation{Department of Physics, Sharif University of Technology, Tehran 14588, Iran}
	
	\author{Sadegh Raeisi \footnote{email: sraeisi@sharif.edu}}
	\affiliation{Department of Physics, Sharif University of Technology, Tehran 14588, Iran}

		\author{Vahid  Karimipour\footnote{email: vahid@sharif.edu, corresponding author} }
	\affiliation{Department of Physics, Sharif University of Technology, Tehran 14588, Iran}

	%\date{\today}
	
	\begin{abstract}
		Having  common reference frames or aligned coordinate systems, is one of the presumptions in witnessing entanglement in a two-party state possessed by  two remote parties. This assumption may fail for many reasons. With an unlimited supply of singlet states, the two parties can first align their coordinate systems and then measure any entanglement witness. In this article, we propose an alternative method which uses the same resource for incomplete teleportation of states between the two parties, enabeling them to witness the entanglement of any shared state by local measurements without the need of prior alignment of the coordinate systems. The method works for any kind of witness and in any dimension.  Beyond the context of Entanglement Witnesses, our method works also for remote measurements of observables of particles (entangled or not) in laboratories which may have deficiency in their resources. \\ \\
	%	PACS numbers: 03.67.Mn, 03.65.Ud, 03.67.−a.
	\end{abstract}

	\maketitle
	
	%%%%%%%%%%%%%%%%%%%%%%%%%%%%%%%%%%%%%%%%%%%%%%%%%%%%%%%%%%%%%%%%

\section{Introduction}

Entanglement is one of the most important features in Quantum mechanics and plays a crucial role in many quantum information protocols, such as quantum key distribution \cite{BB84,SARG04,E91,B92}, quantum teleportation\cite{tele,telePopescu,teleZeilinger}, nonlocality tests\cite{Bell,BellPeres}, quantum computing\cite{qc,qc1}, and quantum secret sharing\cite{qss}. It is important to note that while entangled states are created in a single laboratory, they are to be distributed to parties which are far apart, in some cases even hundreds of kilometers apart. Environmental noise usually deteriorates this entanglement and turns it into a mixed state. For low level of noise, the mixed state may still retain some of the original entanglement to be distilled to maximally entangled states by the two legitimate parties\cite{ent-dist}. Therefore determining whether or not a state is entangled or separable, is a task of utmost importance and has to be done by two remote parties with limited resources, that is, they should do this by Local Operations and Classical Communication (LOCC). In this context, the theory of Entanglement Witnesses (EW) \cite{EntWit}, plays a crucial role. According to this well-developed theory, instead of a full tomography of the state, which is naturally a very costly procedure, one can measure a certain suitably chosen observable to determine whether or not a certain two-party state is entangled. For our purpose, we only need to know the basic concept of Entanglement Witness which we describe below. For many interesting properties and their classification and structure, the reader can consult review articles like \cite{EntWit1,rev-terhal,rev-guhne}. \\

\noindent  Let $H_A$ and $H_B$ be the Hilbert spaces of two particles shared between the two players, called Alice and Bob. For a Hilbert space $H$,  $D(H)$ is the convex set of density matrices related to this Hilbert space, i.e. any $\rho\in D(H)$ is a positive semi-definite operator on $H$ with unit trace. A separable state $\rho$ is of the form $\rho=\sum_i p_i \rho_{a,i}\otimes \rho_{b,i}$, where $p_i$'s are probabilities and $\rho_{a,i}$ and $\rho_{b,i}$ are density matrices. The set of all separable states is obviously a convex set, denoted by $D_{sep}$. 
An Entanglement Witness (EW), is an observable $W$ such that 
\be
Tr(W\rho_{sep})\geq 0\h \forall\rho_{sep}\in D_{sep}, 
\ee
that is, the witness is positive on all the separable states. However an EW is such that it is not positive on all states, i.e. it has negative eigenvalues. Therefore when $Tr(W\rho)<0$, it is a witness that the state $\rho$ is entangled, hence the name.  By now there  have been impressive developments in the theory of Entanglement Witnesses. Besides important concepts of optimality, extremality and decomposability \cite{wit-opt}, many classes of Entanglement Witnesses have been discovered or created \cite{wit1, wit2, wit3, wit4, wit5, wit6, wit7}.  \\

\noindent Naturally an EW is an operator of the form
\be
W=\sum_i A_i\otimes B_i ,
\ee
where Alice and Bob are to measure the observables $A_i$ and $B_i$ on their share of the entangled state and classically communicate the results of these measurements to each other for assembling them to finally determine $Tr(W\rho)$. Obviously this requires that the two players share a common coordinate system in order that this communication have any meaning.  \\

\noindent Although our  motivation is mainly theoretical, there may be some real practical situations where our method is relevant. First of all, due to the rotation of a satellite around the earth and around itself, establishing a Shared Reference Frame (SRF) between earth-satellite or satellite-satellite is a challenging and expensive task
\cite{niskanen}. 
Moreover, some types of noise like unstable fiber communication link or instability in the sending and receiving apparatus is equivalent to the lack of an SRF between sender and receiver. 
Admittedly the two parties can use classical or quantum mechanical methods \cite{gispop,mass,peres1,peres2,reza,bagan,chir,kolen,gold,ours} to align their coordinate systems before making any measurement.   We should note however that the optimal quantum mechanical methods for coordinate system alignment \cite{mass,peres1,peres2,bagan} usually require  $N-$ particle collective measurements which are extremely difficult from   experimental point of view. Alternatively, the two parties can use new methods of quantum communication in the absense of shared reference frames \cite{reza1,reza2,beheshti}. \\

Specifically, we should note a series of works where the parties who do not share a global reference frame, use local random measurement and are able to  generate Bell-violating quantum correlations in bi- and multi-partite pure states \cite{liang1,liang2,liang3,liang4, liang5,liang6}. In these works  the requirement of local calibration of devices was replaced  by  doing measurements in independent bases.  The idea of random measurement to detect entanglement in multi-partite pure states has also been investigated in \cite{liang7} and \cite{liang8}.
 In the spirit of this second line of thought, it is a challenging theoretical question to see, how in the absence of any information about their coordinate systems, 
the two parties can witness the entanglement of an unknown pure or mixed shared state $\rho_{ab}\in D(H_A\otimes H_B)$. We will show that, in any dimension, it is possible to use singlet shared states to witness this entanglement. By singlet states in higher dimensions, we mean states that have zero total angular momentum, see Remark \ref{remark3}).   More precisely, we show that given any Entanglement Witness $W$, the two players who do not know anything about each other's coordinate system, can measure $W$ on the shared state $\rho_{ab}$.  Here and in the rest of the paper, by measurement of observables, we mean measurement the expectation value of them. 
 \\

\noindent The basic idea of the paper is to note the utility of an incomplete teleportation which is based on a specific two-outcome measurement by Alice. We remind that the standard or {\it complete teleportation} protocol of qubits requires a Bell measurement performed by Alice who will send the results by sending two classical bits to Bob. Bob will use this information to fully recover the state hold by Alice. While the explicit calculations are done for pure states, by linearity any mixed state can also be teleported from Alice to Bob. Hence this procedure is capable to send the share of Alice in any entangled state to Bob, a process which is similar to entanglement swapping \cite{tele, ent-swap2} (As shown in figure (\ref{fig1}), a measurement on $a$ and $a'$, changes the pattern of entanglement in the same way as is done in entanglement swapping).  The prerequisite of this protocol is a common coordinate system between the two parties. However, as we will show in the sequel, the crucial point is that to determine whether or not a shared state between Alice and Bob is entangled, a complete teleportation and hence a common coordinate system is not needed. As we will show a specific two-outcome measurement by Alice followed by a one-bit classical communication, can teleport the state in a way which although is incomplete, nevertheless can fully witness the entanglement of the shared state. The nature of this measurement is that no common reference frame is necessary. In fact, we will show that no matter what kind of Entanglement Witness $W$ the two players choose, they can use it without reference to a common reference frame. The only resource that they consume is a singlet state (in the particular dimension of states) and a single bit of classical information. Needless to say, in any type of measurement, including those of Entanglement Witnesses, to reach a reliable statistics of results, a multitude of states is needed which accordingly requires the same number of shared singlet states.  In passing we should note that besides witnessing entanglement, the method presented here, can be used to make any kind of measurement on a shared state between the two parties. At the end, we argue that besides removing the need for common coordinate systems, our method has other applications which may go beyond the simple determination of whether a shared state is entangled or not. \\

\noindent The structure of this paper is as follows: In section (\ref{not}), we describe our special notation, in section (\ref{qubitW}) we explain the procedure for qubits and in section (\ref{quditW}) we generalize it to arbitrary dimensional states. In section (\ref{rm}), we show how this method can be adapted to remote measurement of non-local observables. We conclude with a discussion in section (\ref{con}).

\section{Notatioins and conventions}\label{not}
\begin{figure}[t]
	\centering
	\includegraphics[width=\linewidth]{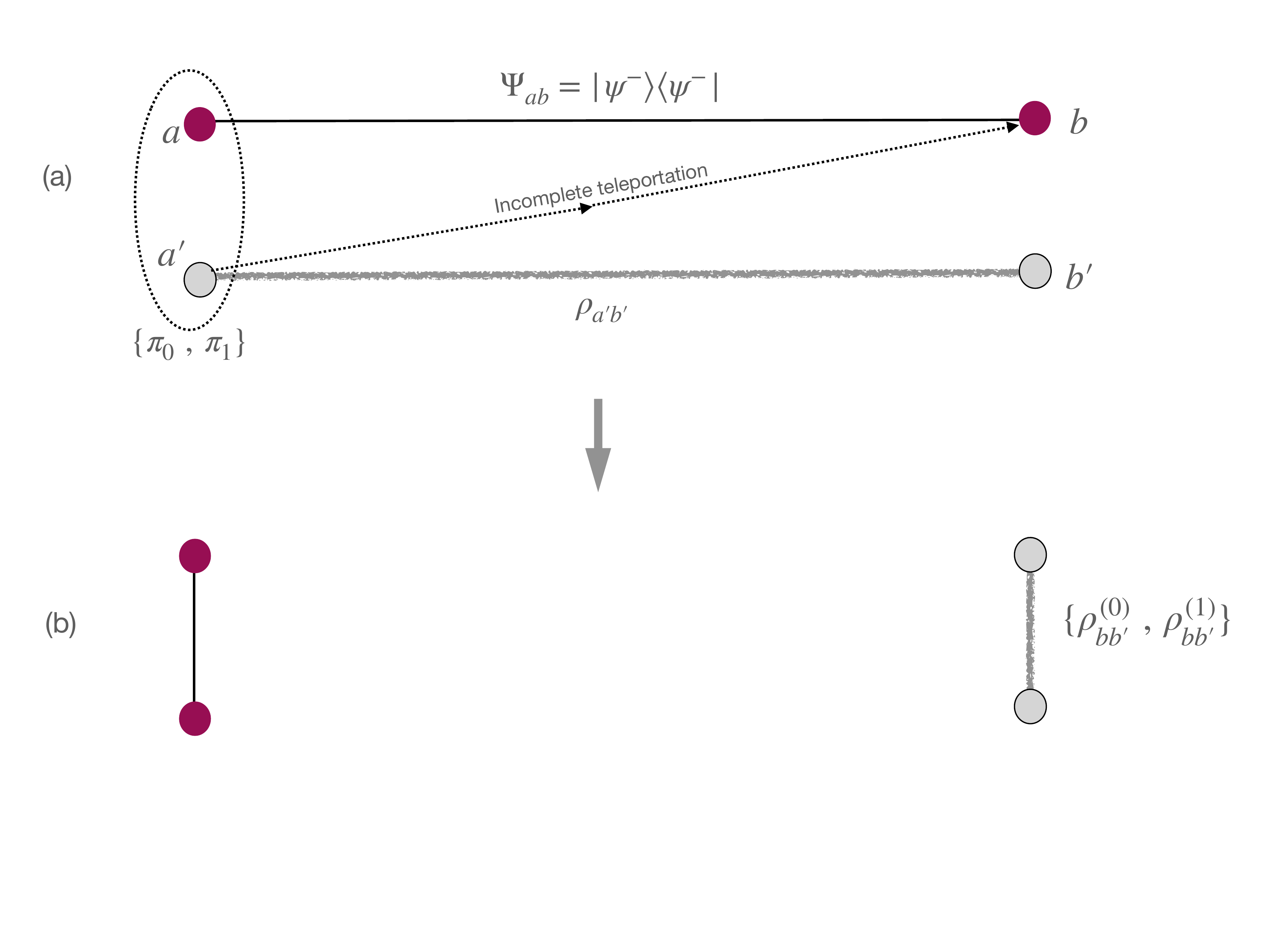}\vspace{-2cm}
	\caption{An unknown state $\rho_{a'b'}$ and a resource state $\psi_{ab}$ shared between Alice and Bob. Alice performs rotation-invariant measurements, i.e. $\pi_0,\pi_1$, and sends her results to Bob. The share labeled by  $a'$ of an uknown state  $\rho_{a'b'}$ is incompletely teleported to Bob, who now possess it in the form $\rho_{bb'}$. Bob can now witness the entanglement of this state which is the entanglement of the original states $\rho_{a'b'}$. }
	\label{fig1}
\end{figure}

Figure (\ref{fig1}) shows the main setup which we use to explain our notations. 
\noindent Taking care of these notations is important for following the arguments. The two players are A (Alice) and B (Bob). Particles in possession of Alice are labeled $a$ and $ a'$, those of Bob by $b$ and $b'$. In any equation, we point to these labels only on the first side of the equation and inside the ket and bra notation. For example, an equation like 
$$
|\psi_{ab}\ra=\frac{1}{\sqrt{2}}(|0,1\ra-|1,0\ra),
$$
(without writing the indices $a$ and $b$ on the right hand side)
indicates that the state of the particles $a$ of Alice and $b$ of Bob is a singlet state, while an equation like 
 $$
 |\psi_{bb'}\ra=\frac{1}{\sqrt{2}}(|0,1\ra-|1,0\ra),
 $$
 indicates that the same state is now in possession of Bob. The partial traces on the left hand side of equations are labeled as $Tr_a$ or $Tr_b$ and those on the right hand side of equations are labeled as $Tr_1$ and $Tr_2$.  
Thus if Alice and Bob share a mixed state $\rho_{ab}$, then we write $Tr_a(\rho_{ab})=\rho_2$ and $Tr_{b}(\rho_{ab})=\rho_1$, where $\rho_1$ and $\rho_2$ are the reduced density matrices of $\rho$.

\subsection{Witnessing entanglement of qubit states}\label{qubitW}
First, we explain in explicit form the incomplete teleportation of qubit states. Let Alice holds a qubit state of the form
\be\label{exp}
|\phi_{a'}\ra=\a |0\ra+\beta |1\ra,
\ee
and let a singlet state  $|\psi\ra$ be shared between Alice and Bob.
\be
|\psi_{ab}\ra=\frac{1}{\sqrt{2}}(|0,1\ra-|1,0\ra).
\ee
Thus Alice holds the qubits $a$ and $a'$ and Bob holds the qubit $b$. The full state is now given by
\be\label{fullstatequbit}
|\Phi_{a'ab}\ra:=|\phi_{a'}\ra\otimes |\psi_{ab}\ra=\frac{1}{\sqrt{2}}(\a |0\ra+\beta |1\ra)\otimes(|0,1\ra-|1,0\ra).
\ee
\begin{remark}\label{remark1}
Here we are writing every state in the frame of Bob, thus all the $|0\ra$'s, and $|1\ra$'s  refer to Bob's coordinate system. In the absence of a shared reference frame, the basis states of Alice, i.e. the eigenstates of the $\sigma_{a,z}$ should be written as $|0'\ra$ and $|1'\ra$ which will be a superposition of $|0\ra$ and $|1\ra$, with unknown coefficients depending on the unknown relative orientation of the two frames. Here, the singlet state can be prepared by Bob or a third party where one of the qubits is then parallel transported to Alice. This assumption of ideal transportation, in a presumably noiseless medium,  has already been used for transferring a direction in a number of works including \cite{gispop,peres1,peres2}.   Obviously the singlet state $|\psi\ra_{ab}$ will not have an expression like $|\psi_{ab}\ra=\frac{1}{\sqrt{2}}(|0',1\ra-|1',0\ra)$, since this requires that the frames of Alice and Bob be indentical which is not the case. 	We also assume that the state $|\phi_{a'}\ra$ which is unknown to Alice,  has  the given expression in equation (\ref{exp}) when written in terms of $|0\ra$ and $|1\ra$. This state if written in the coordinate system of Alice will have different coefficients, but what really matters is the teleportation of the state itself and not these coefficients. As we will see in the sequel, the protocol is based on measurement of expectation values of local observables and for that matter, the protocol needs an unlimited supply of shared entangled states to achieve an unlimited precision. In reality the degree of precision of measurements depend on the number of singlet states which are shared between the two parties.
\end{remark}

Alice now performs a two-outcome measurement to see if the total spin of her two particles is zero or not. The projective measurement that she performs is given by $\{\pi_0,\pi_1\}$, where $\pi_0$ and $\pi_1$ are respectively projectors on the sectors with total spin 0 and 1. Obviously this measurement, being invariant under rotation,  can be performed by Alice, with no need of re-writing them in her coordinate system \cite{relative}. To this end she uses the following relations where in $\psi_{s,m}$, $s$ and $m$ respectively denote the total spin and the $z-$ component of spin.  
\be
\psi_{1,1}=|0,0\ra,\ \ \  \psi_{1,0}=\frac{1}{\sqrt{2}}(|0,1\ra+|1,0\ra),\ \ \ \psi_{1,-1}=|1,1\ra,\ \ 
\ee
and

\be
\ \psi_{0,0}= \frac{1}{\sqrt{2}}(|0,1\ra-|1,0\ra),
\ee
Expanding the full state (\ref{fullstatequbit}), in this new basis, we find
\be
|\Phi_{a'ab}\ra=-\frac{1}{2}\psi_{0,0}\otimes |\phi\ra+\frac{1}{\sqrt{2}}\psi_{1,1}\otimes \sigma^-|\phi\ra- \frac{1}{2}\psi_{1,0}\otimes\sigma_z|\phi\ra -\frac{1}{\sqrt{2}}\psi_{1,-1}\otimes \sigma^+|\phi\ra ,
\ee
where 
$$
\sigma^+=\left(\begin{array}{cc}0&1\\0&0\end{array}\right)\h \sigma^-=\left(\begin{array}{cc}0&0\\ 1&0\end{array}\right).
$$
Therefore with probability $p_0=\frac{1}{4}$, Bob recovers the original state $|\phi\ra$ and with probability $p_1=\frac{3}{4}$, his state will be the following mixed state
\ba
\rho^{(1)}_b&=&\frac{1}{3}\big(2\sigma_-\rho \sigma_++2\sigma_+\rho\sigma_-+\sigma_z\rho\sigma_z\big)\cr
&=&\frac{1}{3}\big(\sigma_x\rho \sigma_x+\sigma_y\rho\sigma_y+\sigma_z\rho\sigma_z\big).
\ea
Using the well-known identity for qubits $\rho+\sum_{a=x,y,z}\sigma_a\rho \sigma_a=2I$, this state can be rewritten as

\be
\rho^{(1)}_b=\frac{2}{3}I-\frac{1}{3}\rho.
\ee
Obviously, Bob cannot extract the original state from this mixed state, hence the name incomplete teleportation. \\

We now consider teleportation of a single qubit which is part of a two-qubit state already shared between Alice and Bob. 
Although we have done the calculations for a pure state, the final results are also valid for arbitrary mixed states $\rho_{a'}$ which Alice may have and hence for arbitrary shared states $\rho_{a'b'}$ between Alice and Bob.
Let us denote the original state by $\rho_{a'b'}$. By incompletely teleporting the qubit $a$ to Bob, as described above, this state will change to either $\rho_{bb'}^{(0)}$ or to $\rho_{bb'}^{(1)}$, depending on the result of Alice's measurement. In our notation, we denote the possessors of states by writing subscripts on the states only in the left hand side of equations and the type of state, without any subscript in the right hand side of equations. Using this convention, we can say that in the two-qubit case,  with probability $p_0=\frac{1}{4}$ this original shared state $\rho_{a'b'}=\rho$ is exactly teleported to become 
\be
\rho^{(0)}_{bb'}=\rho,
\ee 
in possession of Bob and with probability $p_1=\frac{3}{4}$, this state is teleported to become 
\be\label{rho1}
\rho^{(1)}_{bb'}=\frac{2}{3}I\otimes \rho_2-\frac{1}{3}\rho,
\ee
where $\rho_{2}=Tr_1\rho$ is the share of Bob in the original entangled state $\rho_{a'b'}$. While this incomplete scheme succeeds with probability $\frac{1}{4}$ to teleport the state, in the following we show that it will succeed deterministically to witness the entanglement of the shared state.
 The procedure is that Alice informs Bob of the result of her measurement by sending him a classical bit $0$ or $1$. Depending on this classical bit Bob measures a different operator, which are correspondingly denoted by 
$W_{bb'}^{(0)}$ and $W_{bb'}^{(1)}$. Obviously the operator $W_{bb'}^{(0)}$ should be the same as $W$, since in this case, the original state has been teleported to Bob. The second operaotr $W_{bb'}^{(1)}$ should be chosen so that the following equality holds:
\be\label{WW0}
Tr\big(W^{(1)}\rho^{(1)}\big)=Tr(W\rho),
\ee
where for simplicity of notation we have suppressed the indices $b$ and $b'$ from both $W$ and $\rho$ on the left hand side. This equation simply means that the measurements on the two outcomes must be such that they convey the same information about $\rho$ as the original witness $W$ was intended to do.  In view of the form of ($\rho^{(1)}$) as in (\ref{rho1}), and using the identity ($Tr((I\otimes W_2)\rho)=Tr(W(I\otimes \rho_2))$), (where $W_2:=Tr_1(W)$) this equation uniquly determines $W^{(1)}$, which we indicate below together with $W^{(0)}$ for completeness:
\begin{equation}\label{w2cases}
	\begin{cases}
		W^{(0)}_{bb'}=W\\
	W^{(1)}_{bb'}=  2(I\otimes W_2)-3W,
	\end{cases}
\end{equation}
where $W_2=Tr_1W$. The reader can see a proof of this for the general case in equation (\ref{pproof2}).\\

\begin{remark}\label{remark2}
 As we will expalin later, it is important to note that this does not really demand a different experimental setup for each classical bit that Bob receives, rather it only requires a different manipulation of experimental data at the end of the protocol, when all the classical bits have been communicated. The example below illustrates this point.
 \end{remark}

In this way, Alice and Bob can measure the expectation value of the entanglement witness $W$, without the need for a shared coordinate system. This however comes at a cost which is consuming the resource of shared entangled states. 
 
\subsection{An example}

In \cite{ricardi}, a set of optimal Entanglement Witnesses have been constructed which use a minimal set of measurements, namely measurements of diagonally correlated Pauli matrices 

$$
S = \{ \sigma_x \otimes \sigma_x , \sigma_y \otimes \sigma_y , \sigma_z \otimes \sigma_z  \},
$$
(i.e. in contrast to operators of the form $\sigma_i\otimes \sigma_j\ \ i\ne j)$. 
These witnesses are of the form
\be
\tilde{W}=|\phi\ra\la \phi|^\Gamma, 
\ee
where $\Gamma$ denotes partial transpose and $|\phi\ra$ has the following general form
\be
|\phi\ra=a|\phi_1\ra+b|\phi_2\ra ,
\ee
in which $|\phi_1\ra$ and $|\phi_2\ra$, depending on the witness, are two of the Bell states. An  example of this type of Entanglement Witness, when written in terms of Pauli operators takes the following form
\be
\tilde{W}=\frac{1}{4}\big[I\otimes I +\sigma_z\otimes \sigma_z+(a^2-b^2)\big(\sigma_x\otimes \sigma_x+\sigma_y\otimes \sigma_y)+2ab\big(\sigma_z\otimes I+I\otimes \sigma_z\big)\big],
\ee
from which we find $\tilde{W}_2=Tr_1 \tilde{W}=\frac{1}{2}(I+2ab\sigma_z)$ and hence
\begin{equation}
	\begin{cases}
		\tilde{W}^{(0)}_{bb'}=\tilde{W}\\
		\tilde{W}^{(1)}_{bb'}=  I\otimes I+2ab I\otimes \sigma_z-3\tilde{W}.
	\end{cases}
\end{equation}

Note that these two different forms for witnesses, do  not require different experimental setups. In fact, as noted in \cite{ricardi}, the mean values of the one qubit operators $I\otimes \sigma_a$ and $\sigma_a\otimes I$ can be obtained from the mean values of $\sigma_a\otimes \sigma_a$ by simply ignoring the statistics of one side. In fact in both cases, only the diagonally-correlated Pauli operators $\{\sigma_i\otimes \sigma_i, i=x,y,z\}$ are measured, but the data are handled differently depending on the classical bit that Alice sends.  

\subsection{Witnessing entanglement of higher dimensional states}\label{quditW}
We now turn our attention to witnesses for higher dimensional states. The well known protocol for  teleporting a $d-$ dimensional state, is based on sharing a Bell state of the form $|\Phi_{00}\ra=\frac{1}{\sqrt{d}}\sum_{k=0}^{d-1}|k,k\ra$ and making a measurement in the Bell basis 
$
\label{bell}|\Phi_{m,n}\ra=\frac{1}{\sqrt{d}}\sum_{k=0}^{d-1}\omega^{km}|k,k+n\ra ,
$
where $\omega^d=1$. However, the incomplete teleportation which we have described and is possible in the absence of shared reference frames, depends  critically on a specific type of singlet state (a state with zero total angular momentum), being shared between the two parties. 
Such a state which we denote by 
$|\Psi_{ab}\ra$ is invariant under a representation $R\otimes R\in SO(3)\times SO(3)$ and is different from the state $|\phi_{00}\ra$ which is invarint under $U\otimes U^*\in SU(d)\times SU(d)$.   We assume that the states are $d=2j+1$ dimensional where $j$ denotes the total spin of each particle. 
To proceed we first need the general form of a singlet state of two spin $j$ particles. It is easily seen that this unique singlet state is given by
\be\label{sing}
|\Psi_{ab}\ra=\frac{1}{\sqrt{2j+1}}\sum_{m=-j}^j(-1)^m |m_a,-m_b\ra ,
\ee
where in this and later formulas, we have used the abbreviated angular momentum notation $|m\ra$ instead of $|j,m\ra$.  \\

\begin{remark}\label{remark3}
The term singlet  is normally used when we add the angular momentum of two spin $1/2$ particles, where there is only one state with zero total angular momentum. In higher dimensions, there are more than one state with zero total angular momentum. Here we mean a specific type of such state whose explicit form is given in eq. (\ref{sing}).
\end{remark}
One can verify that this state is a singlet state by simply checking that 
\be
J^{tot}_z|\Psi\ra=J^{tot}_{\pm}|\Psi\ra=0,
\ee
where $J^{tot}_z$ and $J^{tot}_\pm$ are the total spin-$j$ represenation of angular momentum operators $J_\mu^{tot}=J_\mu^{(a)}\otimes I^{(b)}+I^{(a)}\otimes J_\mu^{(b)}$ and $\mu$ stands for $z$ and $\pm$. Using abbreviated notation ($|m\ra\equiv |j,m\ra$), the action of angular momentum operators on each part ($a$ or $b$) of the state is given by 
\be
J_z|m\ra=m|m\ra, \ \ \ \ \  J_{\pm}|m\ra=\sqrt{j(j+1)-m(m\pm1)}|m\pm1\ra.
\ee
  The singlet state, having zero total angular momentum, is rotationally invariant and has the same expression in every coordinate system. \\

\begin{remark}\label{remark4}
 Although we use the language of spins, the scheme works for the case where information is encoded in other degrees of freedom, such as polarization of photons, ground and excited states of a two-level atom etc.  In this general setting, a unitary transformation $U$ which up to now we have called rotation, is any unitary operation which transforms the un-primed states into primed states. A singlet state in this context is any state which is invariant under the operation $U\otimes U$.  The measurement operator $\pi_0$ retains its meaning. In fact for  any $d-$ dimensional system, i.e. for qudits, we can set aside the language of spins in so much as the two parties can prepare singlet states and can perform measurements of total spins (in its general meaning defined above). This requires that the two players should be able to use  conventional gates for qudit systems and perform the above tasks. We will show this in the appendix.
\end{remark}

  \noindent Similar to the qubit case, here again, Alice makes a measurement by two projectors $\{\pi_0,\pi_1\}$, but now  $\pi_0$ is the projection on the spin $0$ sector of the two particles and $\pi_1=I-\pi_0$ is the projection on any other spin different from zero.
 Thus, this measurement simply determines whether the total spin of two particles is zero or not and does not determine the total spin of the two particles and this is all that we need for incomplete teleportation. For now, consider the incomplete teleportation of a single particle in the state $\rho_{a'}$ by Alice.  Once we learn this, we can extend it to incomplete teleportation of entangled states, i.e. to incomplete entanglement swapping. The measurement by Alice, is a quantum operation as follows: 
\be
\rho_{a'}\lo p_0\rho^{(0)}_b+p_1\rho^{(1)}_b=:X_b^{(0)}+X_b^{(1)} ,
\ee
where 
\be\label{xib}
X^{(i)}_b=Tr_{aa'}\big[(\pi_i)_{aa'}(\rho_{a'}\otimes \Psi_{ab})(\pi_i)_{aa'}\big]\h i=0,1 ,
\ee
and the probabilities are
\be
p_i=Tr_{aa'b}\big[(\pi_i)_{aa'}(\rho_{a'}\otimes \Psi_{ab})(\pi_i)_{aa'}\big].
\ee
Note that the normalized density matrices are obtained from
\be\label{rhox} \rho^{(i)}_b=\frac{X^{(i)}_b}{p_i}.\ee

The basis states $|k_a\ra$ are normalized and we have
\be\label{norm}
\la k_a|l_a\ra=\delta_{k,l}\h \la k_a|l_b\ra\equiv |l_b\ra\la k_a|,
\ee
where the second equality is due to the two states belonging to two different spaces. From this expression one finds
\be\label{psipsi}
\la \Psi_{aa'}|\Psi_{ab}\ra=\frac{1}{2j+1}\sum_{m=-j}^j |m_b\ra \la m_{a'}|.
\ee
Consider now an arbitrary state of Alice in the form
\be
\rho_{a'}=\sum_{k,l}\rho_{k,l}|k_{a'}\ra\la l_{a'}|.
\ee
From (\ref{xib}), and in view of ($\pi_0=|\Psi\ra\la \Psi|$) we find
\be
X^{(0)}_b=\la\Psi_{aa'}|\big[\rho_{a'}\otimes |\Psi_{ab}\ra\la\Psi_{ab}| \big]|\Psi_{aa'}\ra  ,
\ee
or 
\be X^{(0)}_b=\sum_{k,l}\rho_{kl}\la \Psi_{aa'}|\Psi_{ab}\ra \big[| k_{a'}\ra\la l_{a'}|\big]\la \Psi_{ab}|\Psi_{aa'}\ra.
\ee
Using (\ref{psipsi}), this is expanded to 
\be
X^{(0)}_b=\frac{1}{(2j+1)^2}\sum_{k,l}\rho_{kl}\sum_{n}|n_b\ra\la n_{a'}| \big[| k_{a'}\ra\la l_{a'}|\big]\sum_{m}|m_{a'}\ra\la m_{b}| ,
\ee
which in view of (\ref{norm}), is simplified to
\be\label{x0d}
X^{(0)}_b=\frac{1}{(2j+1)^2}\sum_{k,l}\rho_{kl}|k_b\ra\la l_{b} | =\frac{1}{(2j+1)^2}\rho,
\ee
which shows that the original state of Alice ($\rho_{a'}=\rho$) has been teleported to Bob with a probability  $p_0=\frac{1}{(2j+1)^2}$. To find the output state when the result is $1$, we note that 
\ba\label{x1d}
X^{(1)}_b&=&Tr_{aa'}\big[(\rho_{a'}\otimes \Psi_{ab})(\pi_1)_{aa'}\big]\cr
&=&Tr_{aa'}\big[(\rho_{a'}\otimes \Psi_{ab})(I-\pi_0)_{aa'}\big]\cr
&=&Tr_{aa'}(\rho_{a'}\otimes \Psi_{ab})-\frac{1}{(2j+1)^2}\rho\cr
&=&Tr_{a'}\rho_{a'}Tr_a \Psi_{ab}-\frac{1}{(2j+1)^2}\rho\cr
&=&\frac{1}{2j+1}I-\frac{1}{(2j+1)^2}\rho,
\ea
where in the last line, we have used $Tr(\rho_{a'})=1$ and $Tr_a\Psi_{ab}=\frac{1}{2j+1}I$ and have used our conventions for subscripts denoting the states. 
We also find $Tr(X^{(1)}_b)=1-\frac{1}{(2j+1)^2}=\frac{4j(j+1)}{(2j+1)^2}$. The original state of Alice, $\rho_{a'}$, being a mixed state, allows us to extend these results to teleportation of entangled states or equivalently to entanglement swapping from the pair $(a'b')$ to the pair $(bb')$ as shown in figure (\ref{fig1}). To summarize, equations (\ref{rhox}),  (\ref{x0d}) and (\ref{x1d}) yield the final results of entanglement swapping or teleported states:
\be\label{rrr}
\rho^{(0)}_{bb'}=\rho\ \ \ \ {\rm and} \ \ \ \ \rho^{(1)}_{bb'}=\frac{(2j+1)(I\otimes \rho_2)-\rho}{4j(j+1)},
\ee
which are produced with respective probabilities
\be
p_0=\frac{1}{(2j+1)^2}\ \ \ \ {\rm and}\ \ \ \  p_1=1-\frac{1}{(2j+1)^2}.
\ee
Here $\rho$ is the original entangled states between Alice and Bob, i.e. $\rho_{a'b'}=\rho$.
In exact continuation of the analysis for the qubit case, we now find that the two witnesses are 
\begin{equation}\label{www}
	\begin{cases}
		W^{(0)}_{bb'}=W\\
		W^{(1)}_{bb'}=  (2j+1)I\otimes W_2-4j(j+1)W.
	\end{cases}
\end{equation}
It is in order to provide the explicit calculation for this general case, that is to show that $Tr(W^{(1)}_{bb'}\rho^{(1)}_{bb'})=Tr(W\rho)$. The proof is as follows: from (\ref{www}) and (\ref{rrr}), we find
\be\label{pproof2} Tr(W_{bb'}^{(1)}\rho^{(1)}_{bb'})=\frac{1}{4j(j+1)}Tr\Big[\Big( (2j+1)I\otimes W_2-4j(j+1)W\Big)\Big({(2j+1)(I\otimes \rho_2)-\rho}\Big)\Big].\ee
Using the identities
\be\label{pproof1}
Tr((I\otimes W_2)(I\otimes \rho_2))=(2j+1)Tr_2(W_2\rho_2)\ \ \  {\rm and}\ \ \   Tr((I\otimes W_2)\rho)=Tr(W(I\otimes \rho_2))=Tr_2(W_2\rho_2)
\ee
we find that many of the terms in the right hand side of (\ref{pproof1}) vanish and we are left with $Tr(W\rho)$.\\

In the special case of qubits ($j=\frac{1}{2}$), we recover the results for qubit case (\ref{w2cases}). In this way, any of the higher dimensional Entanglement Witnesses constructed in \cite{ricardi} can be used to detect entanglement of states in remote places, wihtout the need of a shared reference frame.

\section{Remote measurement}\label{rm}
Although our main motivation has been to witness the entanglement of states where the two parties need only check the inequality $Tr(W\rho)$ remotely,  our method can be used to remotely measure non-local observables. Here and in the sequel by {\it measurement}, we mean determining {\it the expectation value of an observable}. This is specially useful when one of the parties does not have sufficient experimental resources to make any type of measurement. In this case, the method presented for entangelement witnesses, can be slightly modified for adapttion  to this new situation. We assume that Alice and Bob want to make a measurement of a non-local observable $G$ on a shared state $\rho_{a'b'}$. Following the same method as explained in previous sections, we find that  $\rho^{(0)}_{bb'}=\rho$ and $\rho^{(1)}_{bb'}=\frac{(2j+1)(I\otimes \rho_2)-\rho}{4j(j+1)}$ with the same probabilities as before and Bob makes local measurements $G^{(0)}_{bb'}$ and $G^{(1)}_{bb'}$ which have the same form as in (\ref{www}) with $W$ replaced with $G$ everywhere:
\begin{equation}
	\begin{cases}
		G^{(0)}_{bb'}=G\\
		G^{(1)}_{bb'}=  (2j+1)I\otimes G_2-4j(j+1)G.
	\end{cases}
\end{equation}
In this way, Alice and Bob, without having a common reference frame, and with different resources for measurements, are able to measure any non-local observable $G$ on their shared state.

\section{Conclusion}\label{con}
We have presented a method for witnessing entanglement of a two-particle state which has been shared between two remote parties, even when the two parties do not have any common coordinate system. The method works for any dimension and for any kind of Entanglement Witness $W$.  Like the case where the two particles are in one single laboratory, measuring an Entanglement Witness requires many copies of the state and hence, it is required that the two parties share a sufficient number of singlet states, possibly sent to them by a  third party, to ascertain a reliable statistics of their measurements. Besides the lack of shared reference frame, there is one more important motivation for our method. The need may sometimes arise when one of the two parties has the necessary complex resources for measuring an arbitrary observable $G$ (not necessarily an Entanglement Witness $W$), on a state which is situated in a remote laboratory (i.e. a satellite) with limited resources  for measurement. In such a case, our method allows the state to be incompletely teleported to the laboratory with sufficient resources for measurement. As mentioned in the introduction, the works reported in  \cite{liang1,liang2,liang3,liang4,liang5,liang5,liang6,liang7,liang8} use random local measurments in specific bases to detect non-classical correlations in pure states, in a (shared reference frame and device)-independent way . Their criteria for detecting non-classical correlation is violation of the Bell inequality. Detecting this violation is obviously more demanding  than measuring a single suitable observable, i.e. an entanglement witness. Moreover, an entanglement witness has the advantage of being sutiable also for detecting entanglement in mixed states.  Therefore it would be interesting to see if the method of random measurements developed in  \cite{liang1,liang2,liang3,liang4,liang5,liang5,liang6,liang7,liang8} can be extended to entanglement witnesses.  

\section{Appendix: Generating singlet states by $d-$ level quantum circuits}
In this appendix our scheme is not specific to spin$-j$ systems and works for any $d-$ level system, i.e. any qudit. We need to show that with the standard gate set of qudit systems, namely the generalized Pauli operators, the Hadamard and the CNOT gate, one can prepare a singlet state and can make a measurement of $\{\pi_0,\pi_1\}$, as described in the text. Needless to say, here we are concerned with theoretical and not experimental aspect of this task. To do this, we first map the angular momentum states of a spin $j-$ particle to the conventional states of a qudit.  Remebering the abbreviated notation for spin states ($|j,m\ra\equiv |m\ra$), this  mapping is as follows:
$$
\{|-j\ra,\ |-j+1\ra, \ \cdots |j-1\ra, |j\ra\} \ \  \ \lo \{ \kett{0},\ \kett{1},\ \cdots \ \kett{d-2},\  \kett{d-1}\} 
$$
where $d=2j+1$ and we have used the $\kett{.}$ notation for the qudit states to distinguish them from the spin states. Therefore the singlet state is identified with the following qudit state of two particles,
\be
|\Psi\ra=\frac{1}{\sqrt{d}}\sum_{m=0}^{d-1}(-1)^m\kett{m, d-1-m}.
\ee 
We will now show that this state can be prepared by a qudit circuit, comprised of  generalized Pauli operators and the Hadamard operator. These operators are defined as follows:
 
\be
X\kett{k}=\kett{k+1}, \h Z\kett{k}=\omega^k\kett{k}, \h \omega^d=1,
\ee
and
\be
H\kett{k}=\frac{1}{\sqrt{d}}\sum_{l=0}^{d-1}\omega^{kl}\kett{l},\h CNOT\kett{k,l}=\kett{k,k+l\ mod \ d}.
\ee
We will later need the curious property of the Hadamard operator, namely 
\be
H^2=\sum_{k=0}^d\kett{-k}\bbra{k}. 
\ee
The cyclic property of all these operators imply that the singlet state which we need to prepare is of the form 
\be
|\Psi\ra=\frac{1}{\sqrt{d}}\sum_{k=0}^{d-1}(-1)^k\kett{k,-1-k}.
\ee
A simple examination shows that the singlet state can be prepared from the product state $\kett{0,0}$ by a simple quantum circuit, namely that 
\be
|\Psi\ra=(Z^{\frac{d}{2}}\otimes X^{-1}H^2)CNOT(H\otimes I)\kett{0,0},
\ee
the verification of which is straightforward. Conversely, as in the qubit case, if the reverse circuit $(H\otimes I)CNOT(Z^{\frac{d}{2}}\otimes X^{-1}H^2)$, projects any arbitrary state $\kett{\Phi}$  onto $\kett{0,0}$, it means that the total spin of the state $\kett{\Phi}$ is zero. Any other outcome, indicates a total spin different from zero. This stems from the orthogonality of the spin states and the unitarity of the circuit.


\begin{thebibliography}{}

\bibitem{BB84} C. H. Bennett and G. Brassard, in \emph{Quantum cryptography: Public key distribution and coin tossing}, IEEE International Conference on Computers, Systems, and Signal Processing (India, 1984).

\bibitem{SARG04} C. Branciard, N. Gisin, B. Kraus, V. Scarani, \emph{Security of two quantum cryptography protocols using the same four qubit states}, Phys. Rev. A 72 (3): 032301 (2004).

\bibitem{E91} A. K. Ekert, \emph{Quantum cryptography based on Bell's theorem}, Phys. Rev. Lett. 67 (6): 661–663 (1991).

\bibitem{B92} C. H. Bennett, \emph{Quantum cryptography using any two nonorthogonal states}, Phys. Rev. Lett. 68.3121 (1992).

\bibitem{tele} C. H. Bennett, G. Brassard, R. Crépeau, R. Jozsa, A. Peres, W. K. Wootters, \emph{Teleporting an unknown quantum state via dual classical and Einstein-Podolsky-Rosen channels}, Phys. Rev. Lett. 70.1895 (1993).

\bibitem{telePopescu} D. Boschi, S. Branca, F. De Martini, L. Hardy, S. Popescu, \emph{Experimental Realization of Teleporting an Unknown Pure Quantum State via Dual Classical and Einstein-Podolsky-Rosen Channels}, Phys. Rev. Lett. 80 (6) 1121–1125 (1998).

\bibitem{teleZeilinger} D. Bouwmeester, J. W. Pan, K. Mattle, M. Eibl, H. Weinfurter, A. Zeilinger, \emph{Experimental quantum teleportation}, Nature. 390 (6660) 575–579 (1997).

\bibitem{Bell} J. S. Bell, \emph{On the Einstein Podolsky Rosen Paradox}, Physics Physique 1 (3) 195–200 (1964).

\bibitem{BellPeres} A. Peres, \emph{All the Bell inequalities}, Foundations of Physics 29 (1999) 589-614 (1999).

\bibitem{qc} R. Jozsa, \emph{Entanglement and quantum computation}, quant-ph/9707034 (1997).

\bibitem{qc1} T. D. Ladd, F. Jelezko, R. Laflamme, Y. Nakamura, C. Monroe, and J. L. O’Brien, \emph{Quantum Computing}, Nature (London) 464, 45 (2010).

\bibitem{qss} M. Hillery, V. Buzek and A. Berthiaume, \emph{Quantum secret sharing}, Phys. Rev. A, 59, 1829 (1999).

\bibitem{ent-dist} C. H. Bennett, G. Brassard, S. Popescu, B. Schumacher, J. A. Smolin, W. K. Wootters, \emph{Purification of Noisy Entanglement and Faithful Teleportation via Noisy Channels}, Phys.Rev.Lett.76:722-725,1996.

\bibitem{EntWit} M. Horodecki, P. Horodecki, R. Horodecki, \emph{Separability of mixed states: necessary and sufficient conditions}, Physics Letters A 223 (1996).

\bibitem{EntWit1} D. Chru\'{s}ci\'{n}ski and G. Sarbicki, \emph{Entanglement witnesses: construction, analysis and classification}, J. Phys. A: Math. Theor. 47, 483001 (2014).


\bibitem{rev-terhal} B. M. Terhal, \emph{Detecting Quantum Entanglement}, Journal of Theoretical Computer Science 287(1), 313-335 (2002).

\bibitem{rev-guhne} O. G\"{u}hne, G. Toth, \emph{Entanglement detection}, Physics Reports 474, 1 (2009).



\bibitem{wit-opt} M. Lewenstein, B. Kraus, J. I. Cirac, and P. Horodecki, \emph{Optimization of entanglement witnesses}, Phys. Rev. A 62, 052310 (2000).


\bibitem{wit1} A. Sanpera, D. Bruss, and M. Lewenstein, \emph{Schmidt number witnesses and bound entanglement}, Phys. Rev. A 63, 050301 (2001).

\bibitem{wit2} M. Marciniak, \emph{Rank properties of exposed positive maps}, Linear and Multilinear Algebra, vol. 61, 2013, pp. 970-975.

\bibitem{wit3} D. Chruściński, A. Kossakowski, \emph{On the structure of entanglement witnesses and new class of positive indecomposable maps}, Open Systems and Inf. Dynamics, 14, 275 (2007).

\bibitem{wit4} L. O. Hansen, A. Hauge, J. Myrheim, P. Ø. Sollid, \emph{Extremal entanglement witnesses}, International Journal of Quantum Information, Vol. 13, No. 08, 1550060 (2015).

\bibitem{wit5} M. Gluza, M. Kliesch, J. Eisert, L. Aolita, \emph{Fidelity witnesses for fermionic quantum simulations}, Phys. Rev. Lett. 120, 190501 (2018).

\bibitem{wit6} C. Branciard, D. Rosset, Y. Liang, N. Gisin, \emph{Measurement-Device-Independent Entanglement Witnesses for All Entangled Quantum States}, Phys. Rev. Lett. 110, 060405 (2013).

\bibitem{wit7} J. Bae, D. Chruściński, B. C. Hiesmayr, \emph{Mirrored entanglement witnesses}, NPJ Quantum Inf 6, 15 (2020).


\bibitem{niskanen} J. Wabnig, D. Bitauld, H. W. Li, A. Laing, J. L. O'Brien, A. O. Niskanen, \emph{Demonstration of Free-space Reference Frame Independent Quantum Key Distribution}, New J. Phys. 15 073001 (2013).



\bibitem{gispop} N. Gisin and S. Popescu, \emph{Spin Flips and Quantum Information for Antiparallel Spins}, Phys. Rev. Lett. \textbf{83}, 432 (1999).
	
	
	\bibitem{mass} S. Massar and S. Popescu, \emph{Optimal Extraction of Information from Finite Quantum Ensembles}, Phys. Rev. Lett. \textbf{74}, 1259 (1995).
	
	
\bibitem{peres1} A. Peres and P. F. Scudo, \emph{Entangled Quantum States as Direction Indicators}, Phys. Rev. Lett. 86 (2001) 4160.

\bibitem{peres2} A. Peres and P. F. Scudo, \emph{Transmission of a Cartesian Frame by a Quantum System}, Phys. Rev. Lett. 87 (2001) 167901.

	
\bibitem{reza} F. Rezazadeh, A. Mani, V. Karimipour, \emph{Secure alignment of coordinate systems by using quantum correlation}, Physical Review A 96 (2), 022310.

\bibitem{bagan} E. Bagan, M. Baig, and R. Mun\~{o}z Tapia, \emph{Aligning Reference Frames with Quantum States}, Phys. Rev. Lett. 87, 257903 (2001).

\bibitem{chir}G. Chiribella, G. M. D'Ariano, P. Perinotti  and M. E. Sacchi, \emph{Efficient Use of Quantum Resources for the Transmission of a Reference Frame}, Phys. Rev. Lett.  {\bf 93}, 180503 (2004).

\bibitem{kolen} P. Kolenderski and R. Demkowicz-Dobrzanski, \emph{Optimal state for keeping reference frames aligned and the platonic solids}, Phys. Rev. A 78, 052333 (2008).

\bibitem{gold} A. Z. Goldberg and D. F. V. James, \emph{Quantum-limited Euler angle measurements using anticoherent states} Phys. Rev. A 98, 032113 (2018).	

\bibitem{ours} M. M. R. Koochakie, V. Jannesary, V. Karimipour, \emph{GHZ states as near-optimal states for reference frame alignment}, Quantum Information Processing volume 20, Article number: 329 (2021).
\bibitem{reza1}F Rezazadeh, A Mani, V Karimipour, Quantum key distribution with no shared reference frame,
Quantum Information Processing 19, 1-12 (2020).
\bibitem{reza2}F Rezazadeh, A Mani, V Karimipour, Power of a shared singlet state in comparison to a shared reference frame
Physical Review A 100 (2), 022329 (2019).
\bibitem{beheshti} A Beheshti, S Raeisi, V Karimipour, Entanglement-assisted communication in the absence of shared reference frame, Physical Review A 99 (4), 042330 (2019).


\bibitem{liang1}Yeong-Cherng Liang, Nicholas Harrigan, Stephen D. Bartlett, Terry Rudolph,  Nonclassical correlations from randomly chosen local measurements, 	Physical Review Letters,  104,  050401 (2010).

\bibitem{liang2} Joel J. Wallman, Yeong-Cherng Liang, Stephen D. Bartlett, Generating nonclassical correlations without fully aligning measurements, Phys. Rev. A 83, 022110 (2011).

\bibitem{liang3} Gelo Noel M. Tabia, Varun Satya Raj Bavana, Shih-Xian Yang, Yeong-Cherng Liang, Bell inequality violations with random mutually unbiased bases, Phys. Rev. A 106, 012209 (2022).


\bibitem{liang4} Minh Cong Tran, Borivoje Dakić, François Arnault, Wiesław Laskowski, and Tomasz Paterek, Quantum entanglement from random measurements, Phys. Rev. A 92, 050301(R)  (2015).


\bibitem{liang5} Peter Shadbolt, Tamas Vertesi, Yeong-Cherng Liang, Cyril Branciard, Nicolas Brunner, Jeremy L. O'Brien, Guaranteed violation of a Bell inequality without aligned reference frames or calibrated devices, Scientific Reports 2, 470 (2012).


\bibitem{liang6} Shih-Xian Yang, Gelo Noel Tabia, Pei-Sheng Lin, Yeong-Cherng Liang, Device-independent certification of multipartite entanglement using measurements performed in randomly chosen triads, 	Phys. Rev. A 102, 022419 (2020).


\bibitem{liang7} Andreas Ketterer, Nikolai Wyderka, and Otfried Gühne, Characterizing Multipartite Entanglement with Moments of Random Correlations, Phys. Rev. Lett. 122, 120505  (2019).


\bibitem{liang8} Lukas Knips, Jan Dziewior, Waldemar Kłobus, Wiesław Laskowski, Tomasz Paterek, Peter J. Shadbolt, Harald Weinfurter and Jasmin D. A. Meinecke , Multipartite entanglement analysis from random correlations, npj Quantum Information, 6, 51 (2020).



\bibitem{ent-swap2} M. Żukowski, A. Zeilinger, M. A. Horne, and A. K. Ekert, \emph{‘‘Event-ready-detectors’’ Bell experiment via entanglement}, Phys. Rev. Lett. 71 4287 (1993).

\bibitem{relative}  S. D. Bartlett, T. Rudolph and  R. W. Spekkens,  \emph{Optimal measurements for relative quantum information}  Phys. Rev. A. 70, 032321 (2004).


\bibitem{ricardi} A. Riccardi, D. Chru\'{s}ci\'{n}ski, and C. Macchiavello, \emph{Optimal entanglement witnesses from limited local measurements}, Phys. Rev. A 101, 062319 (2020).






\end{thebibliography}
\end{document}